\documentclass[twocolumn,nodate,showpacs,preprintnumbers,nofootinbib,amsmath,amssymb,aps,prd]{revtex4}

\usepackage{setspace}
\usepackage{amssymb}
\usepackage{graphicx,wrapfig}
\usepackage{enumerate} 
\usepackage{color}
\usepackage{mathrsfs}
\usepackage{subfigure}
\definecolor{mg}{rgb}{0.0, 0.5, 0.0}

\def\be{\nopagebreak[3]\begin{equation}}
\def\ee{\end{equation}}
\def\ba{\nopagebreak[3]\begin{eqnarray}}
\def\ea{\end{eqnarray}}
\newcommand{\f}{\frac}

\def\d{{\rm d}}

\def\t{\tilde}
\def\h{\hat}

\def\db{\delta_b}
\def\dc{\delta_c}
\def\T{\mathcal{T}}

\begin{document}

\title{Quantum Transfiguration of Kruskal Black Holes}
\author{Abhay Ashtekar$^{1}$, Javier Olmedo$^{1}$, Parampreet Singh$^{2}$}
\affiliation {
1. Institute for Gravitation and the Cosmos \& Physics Department, Penn State University,
University Park, PA 16801 \\
2. Department of Physics and Astronomy, Louisiana State University,
Baton Rouge, LA 70803}

\begin{abstract}
%A new effective theory  of macroscopic Kruskal black holes is presented, encompassing
 We present a new effective description of macroscopic Kruskal black holes that  incorporates corrections due to quantum geometry effects of loop quantum gravity. It encompasses both the `interior' region that contains classical singularities and the `exterior' asymptotic region. Singularities  are naturally resolved by the quantum geometry effects of loop quantum gravity, and the resulting quantum extension of the full Kruskal space-time is free of all the known limitations of previous investigations \cite{ab,lm,ck,bv,dc,cgp,cs,oss,cctr,yks,js} of the Schwarzschild interior. We compare and contrast our results with these investigations and also with the expectations based on the AdS/CFT duality \cite{eh}.
\end{abstract}

\pacs{04.60Pp,04.70.Dy,04.20.Dw}
\maketitle

Black hole singularities have drawn great deal of attention in the quantum gravity literature especially over the past two decades (see, e.g., \cite{ab,lm,ck,bv,dc,cgp,cs,oss,cctr,yks,js,
dc2,bkd,eh,gop,crfv,hhcr,djs,mc,bcdhr,apbook}). While there is  general consensus that  these singularities are windows onto physics beyond Einstein, there is still no agreement on how they  are to be  resolved in quantum gravity, and indeed, on whether they would be resolved. For example,  in the Penrose diagram of an evaporating black hole that Hawking drew over 40 years ago \cite{swh}, the singularity persists as part of the future boundary of space-time even after the black hole has completely disappeared. Although this persistence is not based on a detailed calculation, this paradigm  is still widely used.

The goal of this letter is to address this general issue using a new effective theory that describes the quantum corrected geometry of \emph{macroscopic} Schwarzschild-Kruskal black holes.  Salient features of this geometry can be summarized as follows: (i) All curvature scalars have absolute (i.e., mass independent) upper bounds; (ii) Space-time admits an infinite number of trapped, anti-trapped and asymptotic regions;  (iii) In the large mass limit, consecutive  asymptotic regions of the extension have the same ADM mass;  and, (iv) In the low curvature regions (e.g., near and outside  horizons) quantum effects are negligible.  As we discuss below, previous effective theories \cite{ab,lm,bv,dc,cgp,cs,oss,cctr,bkd,djs}  that also resolved the  black hole singularity have some undesirable features. The new description is free of these limitations. 

We will begin with a discussion of our effective dynamics in the phase space, then summarize its  predictions for space-time geometry, and finally compare and contrast our results with those in the literature.

\emph{Phase space of the Schwarzschild `interior':}  As is well-known, the Schwarzschild `interior' --the region of Kruskal space-time bounded by horizons-- is isometric with the vacuum Kantowski-Sachs space-time. It is foliated by spatially homogeneous 3-manifolds $\Sigma$ with topology $\mathbb{R}\times \mathbb{S}^{2}$. Denote the natural coordinates adapted to the spatial isometries by $x\,,\theta,\phi$. Because $\Sigma$ is non-compact in the $x$-direction and fields are homogeneous, in the phase space framework one encounters infinities. Therefore, as is common, we introduce a fiducial cell $\mathcal{C}$ in $\Sigma$ with the same topology $\mathbb{R}\times \mathbb{S}^{2}$ but with $x \in (0, L_{o})$ and restrict phase space variables to $\mathcal{C}$. (Final physical results, of course, have to be well-defined as this 
infrared cut-off $L_{o}$ is removed.) %independent of this infrared cut-off $L_{o}$.) 
In  loop quantum gravity (LQG), they are the gravitational {\rm SU(2)} connections $A_{a}^{i}$ and their canonical conjugate momenta $E^{a}_{i}$  (that represent (densitized) orthonormal triads). Because of the underlying symmetries, the pairs $A_{a}^{i}, E^{a}_{i}$ have the form \cite{ab,cs,aos}
\ba
A^i_a \tau_i \d x^a \hskip-0.1cm = \f{c}{L_{o}}  \tau_3 \d x +  b \, \tau_2 \d \theta 
- b\, \tau_1 \sin \theta \d \phi + \tau_3 \cos \theta \d \phi  %\,\,\nonumber\\
\label{connection}\\
\hskip-0.3cm E^a_i \tau^i \partial_{a} \hskip-0.1cm =  p_c \tau_3 \, \sin \theta 
\, \partial_x + \f{p_b}{ L_o} \tau_2 \sin \theta \, 
\partial_\theta - \f{p_b}{ L_o}  \tau_1 \partial_\phi\,\, \label{triad}
\ea
%
%\ba
%A^i_a \tau_i \d x^a \hskip-0.1cm = (c/L_{o})  \tau_3 \d x +  b \, \tau_2 \d \theta 
%- b\, \tau_1 \sin \theta \d \phi + \tau_3 \cos \theta \d \phi  %\,\,\nonumber\\
%\label{connection}\\
%\hskip-0.3cm E^a_i \tau^i \partial_{a} \hskip-0.1cm =  p_c \tau_3 \, \sin \theta 
%\, \partial_x + ({p_b}/{ L_o}) \tau_2 \sin \theta \, 
%\partial_\theta - ({p_b}/{ L_o})  \tau_1 \partial_\phi\,\, \label{triad}
%\ea
%
where $\tau_{i}$ are the {\rm SU(2)} generators, and  $b,p_{b}$;$\,c,p_{c}$ now represent the canonically conjugate pairs. The Poisson brackets are given by:  $\{c,\,p_c\} \, = \,2 G \gamma$ and  $\{b,\,p_b\} \, = \,  G \gamma$, where $\gamma$ is the Barbero-Immirzi parameter of LQG \cite{apbook}. Given any choice of the time coordinate $\tau$ and the associated lapse $N_{\tau}$, each point in the phase space defines a 4-metric with  Kantowski-Sachs isometries:
\be\label{metric}
 \d s^2 = - N_{\tau}^2 \d \tau^2 + \f{p_b^2}{|p_c| L_o^2} \d x^2 + |p_c| (\d \theta^2 + \sin^2\theta \d \phi^2). 
\ee
Finally, the requirement that physical quantities must be insensitive to rescalings of the fiducial $L_{o}$ implies that  they can depend only on $b, p_{c}$ and the combinations $c/L_{o}$ and $p_{b}/L_{o}$. If one uses the Hamiltonian constraint $H(N)$  of general relativity (GR), as one would expect, the dynamical trajectories on phase space reproduce the Schwarzschild `interior'  geometry \cite{ab,cs,aos}.

\emph{Effective dynamics:} In loop quantum cosmology (LQC),  the full quantum evolution is extremely well approximated by certain quantum corrected, `effective equations'.  For the Schwarzschild interior, we will only consider the analogous effective theory because the explicit action of our quantum constraint operator remains too complicated to explore full quantum dynamics.

The expression of $H[N]$ of the Hamiltonian constraint of LQG contains curvature $F_{ab}^{i}$ of the gravitational connection $A_{a}^{i}$.  In GR, curvature components can be obtained by first considering ratios  $(h_{\Box}/\mathcal{A}_{\Box})$, where $h_{\Box}$ is the holonomy of $A_{a}^{i}$ around suitable plaquettes $\Box$ and $\mathcal{A}_{\Box}$ is the area enclosed by $\Box$, and then taking the limit as the plaquettes shrink to a point. In LQG, the area operator has minimum non-zero eigenvalue $\Delta $=$ 4 \sqrt{3} \pi \gamma \ell_{\rm{Pl}}^2$
--called the \emph{area gap}-- and curvature operators are given by $\h{h}_{\Box}$ where now the plaquettes $\Box$ enclose area $\Delta$ {\cite{apbook,asrev}}.  In our case we need to introduce  three plaquettes  to obtain curvature operators $\h{F}_{\theta,\phi}\,,  \h{F}_{x,\theta}, \,  \h{F}_{x,\phi}$. Lengths of the links in these plaquettes introduce two `quantum parameters', $\dc$ for the $x$-directional link, and $\db$ for links in the 2-spheres  (`quantum' because they depend on the area gap $\Delta$). These parameters feature in the expression of the Hamiltonian constraint and hence also in the dynamical equations in the effective theory. It turns out that the equations can be solved exactly for a convenient choice of the lapse $N$,
\be \label{N}  N = \frac{\gamma \,\mathrm{sgn} \,(p_c) \,|p_c|^{1/2} \,\,\delta_b}{\sin(\db b)},\ee
provided the quantum parameters $\db,\dc$ are chosen to be  appropriate Dirac observables  (i.e. certain phase space functions which are constants of motion in the effective theory).  This is a subtle point because whether a phase space function is a Dirac observable depends on $H(N)$  which itself depends on $\db,\dc$. However, self-consistent choices  of $\db,\dc$ can be made  in a systematic fashion \cite{aos}. Once such choice is made, explicit solutions are given by:
\ba \label{eq:c}
&\tan \Big(\delta_{c}\, c(T) / {2} \Big) =  \mp \f{\gamma L_o \dc}{8 m}\, e^{-2 T} ,\nonumber\\
\label{eq:pc} 
&p_c(T) = 4 m^2 \Big(e^{2 T} + \f{\gamma^2 L_o^2 \dc^2}{64 m^2} e^{-2 T}\Big) , \nonumber\\
\label{eq:b}
&\cos \big(\delta_{b }\,b(T)\big) = b_o \tanh\left(\f{1}{2}\Big(b_o T + 2 \tanh^{-1}\big(\frac{1}{b_o}\big)\Big)\right) ,\nonumber\\
\label{eq:pb}
&p_b(T) = \nonumber - 2m {\gamma L_{o}\, \db \sin(\delta_b b(T))}/( {\sin^{2} (\db b(T)) + \gamma^{2} \db^{2}}). \nonumber
\ea
Here $T$ is the time coordinate defined by $N$, $b_o = (1 + \gamma^2 \db^2)^{1/2}$, and  $m:= (p_{c}\,\sin \dc c)/(\gamma L_{o} \dc)$ is a Dirac observable. A detailed examination shows that $m=GM$ where $M$ is the black hole mass \cite{cs,aos}.
\nopagebreak[3]\begin{figure}% [h] %\bfig
\vskip0.5cm
\includegraphics[width=170pt,height=200pt]{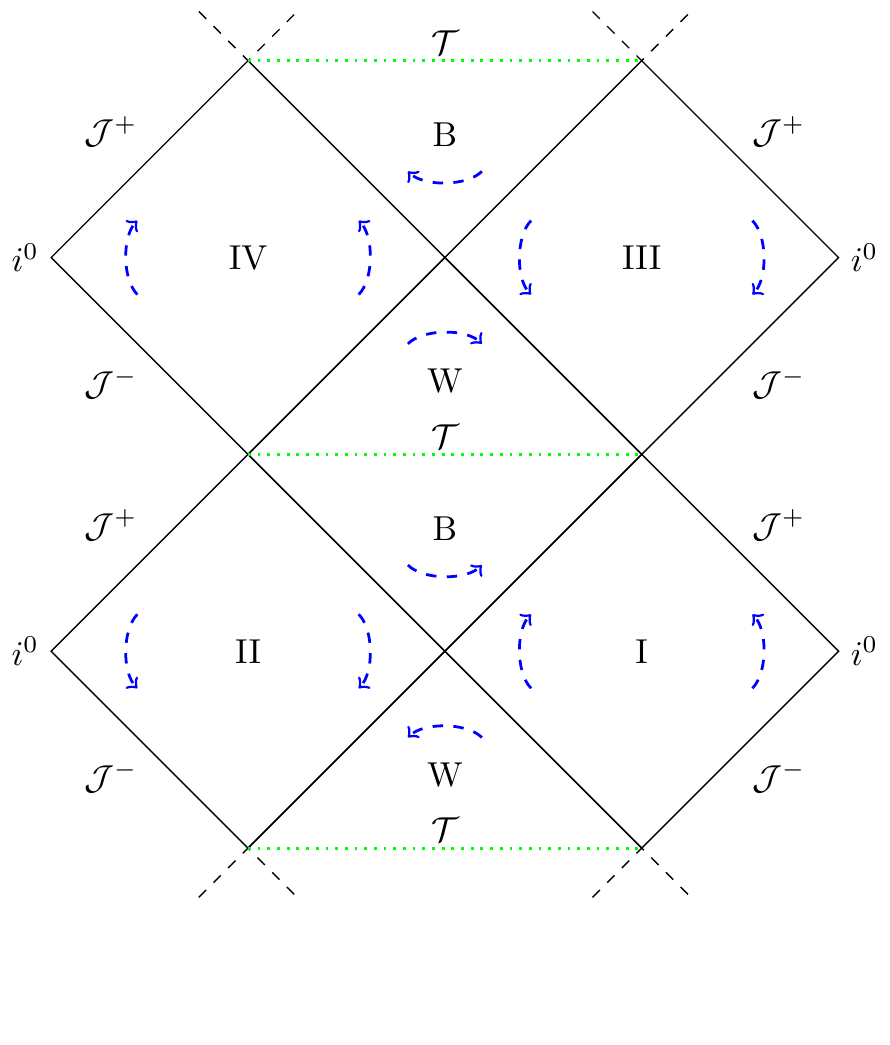}
\vskip-1cm
\caption{\footnotesize{The Penrose digram of the extended Kruskal space-time. The central diamond $B\cup W$ is the `interior', containing the trapped region $B$ and an anti-trapped region $W$, separated by the transition surface $\T$ that replaces the classical singularity.  I, II, III and IV are asymptotic regions and the arrows represent the translational Killing vector $X^{a}\partial_{a} \equiv \partial/\partial x$.}}
\label{fig:4}
\end{figure}

Let us note a few properties of  these quantum corrected dynamical trajectories. The black hole horizon corresponds to $T=0$ where $b$ --and hence also $p_{b}$-- vanishes,  
 and the Killing vector $\partial/\partial x$ becomes null.  $T$  then \emph{decreases} as we move to the future in the interior region. One can calculate expansions $\theta_{\pm}(T)$ of the two null normals $\ell^{a}_{\pm}$ to the round 2-spheres.  As expected, $\theta_{\pm}$ are both negative near the black hole horizon, whence we have a trapped region. However, they both vanish (simultaneously) when $\d p_{c}/\d T$ vanishes. This does occur along each dynamical trajectory, and \emph{occurs once and only once,} at  time $T_{\T}  = (1/2) \ln(\gamma L_{o}\dc/8 m)$. In the space-time picture,  at $T=T_{\T}$ we have a \emph{transition surface} $\T$, to the past of which we have a trapped region and to the future of which we have an anti-trapped region (in which both expansions are positive). Just as the trapped region has a past boundary given by the black hole horizon, the anti-trapped region has a future boundary at a white hole type horizon where, again, $b$ and hence $p_{b}$ vanishes. This occurs at  a finite value of time, $T_{0} := - (4/b_{o})\, \tanh^{-1} (1/b_{o})$.  What happens in the classical limit ($\Delta \to 0$ and hence) $\db\to 0$ and $\dc\to 0$? In this limit,  $T_{0} \to -\infty$ whence $b,c$ diverge and $p_b$, $p_{c}$ vanish; this is just the classical singularity. In this precise sense, \emph{the transition surface $\T$ of the effective theory replaces the singularity of GR.} While in GR  the `interior' region is bounded by the black hole horizon in the past and the singularity in the future,  in the quantum corrected theory it is bounded by the black hole horizon in the past ($T$=0) and a white-hole horizon in the future ($T$=$T_{0}$) --depicted by the central diamond in Fig. \ref{fig:4}.\smallskip

\emph{Quantum parameters:} First investigations of the Schwarzschild `interior' \cite{ab,lm,cgp}  used the so-called  `$\mu_{o}$-scheme' that had been introduced earlier in Friedmann, Lemaitre, Robertson, Walker (FLRW) cosmology \cite{abl}, and set $\db,\dc$ to a constant, $\delta$. This scheme turned out to have serious limitations in FLRW models and was replaced by the so-called  `$\bar\mu$ scheme'  in \cite{aps3,asrev}. Subsequently,  $\bar{\mu}$-inspired strategies were implemented  for the Schwarzschild `interior' \cite{bv,dc,cctr}. However, detailed investigations showed that both of these schemes lead to  physically undesirable results  in this case \cite{cs,js}. Therefore, we place ourselves `in between' the two strategies and  systematically arrive at the expressions of  $\db,\dc$ by specifying the plaquettes  enclosing minimum area: The plaquettes are now chosen to lie on the transition surface $\T,$ where, as we will see,  curvature invariants assume largest values.  Because each solution admits one and only one transition surface $\T,$\, $\db, \dc$ are now (judiciously chosen) Dirac observables.  They are not constants on the full phase space as in the  `$\mu_{o}$-scheme'  but,  in contrast to the `$\bar\mu$-scheme', they are constant along each effective dynamical trajectory.
%choose $\db,\dc$ judiciously to be certain Dirac observables.  Their expressions are now arrived at systematically by specifying the plaquettes  enclosing minimum area: They are chosen to lie on the transition surface $\T$, where, as we will see,  curvature invariants assume largest values.  Because each solution admits one and only one transition surface $\T$,\, $\db, \dc$ are now Dirac observables. 
%--they will not be constants on the full phase space as in \cite{ab,lm,cgp}, and in contrast to \cite{bv,dc,cctr}, they will  be constant along every effective dynamical trajectory.  {\color{blue} However, while in  forms of $\db,\dc$ are postulated, now they are arrived at systematically by specifying the plaquettes  enclosing minimum area: They are chosen to lie on the transition surface $\T$, where, as we will see,  curvature invariants assume largest values. Because each solution admits one and only one transition surface $\T$,\, $\db, \dc$ are Dirac observables.  
As Ref. \cite{aos} shows in detail, in the large $m$ limit this procedure yields
\be\label{db-dc}
\db=\Big(\frac{\sqrt{\Delta}}{\sqrt{2\pi}\gamma^2m}\Big)^{1/3}, \qquad 
L_{o}\dc=\frac{1}{2} \Big(\frac{\gamma\Delta^2}{4\pi^2 m}\Big)^{1/3}.
\ee
This specific choice plays a key role in freeing our effective dynamics from limitations of previous works. 

\emph{The Schwarzschild  `exterior':} Analysis of the `interior' makes crucial use of spatial homogeneity. Since the `exterior' region does not admit spatially homogeneous slices,  effective theories had not been extended to the `exterior' asymptotic regions.  Note, however, that the asymptotic region \emph{can be} foliated by \emph{time-like} homogeneous slices ($r$=const. in the Schwarzschild coordinates), whence one can construct a Hamiltonian framework based on them. This seems unusual at first but the `evolution' in the radial direction is again generated by a Hamiltonian constraint, equations of motion are again ODEs and they can again be explicitly solved using a convenient choice of the lapse function. (Indeed, one can extend the Hamiltonian framework to incorporate time-like hypersurfaces even for full GR and it would be of interest to extend LQG ideas to include these frameworks.)

Indeed, this Hamiltonian theory can be obtained rather easily  from that of the interior region. Since the homogeneous slices $T$=const. (i.e., $r$=const. in Schwarzschild coordinates) are now time-like, the LQG phase space variables $\t{A}_{a}, \t{E}^{a}$ now take values in ${\rm SU(1,1)}$ rather than ${\rm SU(2)}$. Hence one only needs to replace the ${\rm su(2)}$ basis $\tau_{i}$ by the ${\rm su(1,1)}$ basis $\t\tau_{i}$.  As a result, the connection and the triad now have the same form as (\ref{connection}) and  (\ref{triad}), with replacements $\tau_{i} \to \t\tau_{i}$ and $(b,c;\, p_{b}, p_{c}) \to (\t{b}, \t{c};  \t{p}_{b}, \t{p}_{c})$. Finally, since $\t\tau_{i}$ are given by 
$\t\tau_{1} = i \tau_{1}, \quad \t\tau_{2} = i \tau_{2}, \quad \t\tau_{3} =  \tau_{3}$,  
the phase space description for the `exterior' region can be obtained simply by substitutions
\be \label{substitutions} b \to i\t{b},   \, p_{b} \to i \t{p}_{b}; \qquad c\to  \t{c},\, p_{c} \to \t{p}_{c}, \ee
in the phase space description of the `interior', with real $\t{b}, \t{p}_{b}, \t{b}, \t{p}_{c}$  
(for details, see \cite{aos}). Then the dynamical trajectories for the `exterior' region are given by:
\hskip-0.3cm
\ba 
 \label{eq:ct} 
&\tan \Big(\f{\delta_{\t{c}}\, \t{c}(T)}{2} \Big)=  \mp \f{\gamma L_o \delta_{\t{c}}}{8 m} e^{-2 T}, \nonumber\\
\label{eq:pct} 
&\t{p}_c(T) = 4 m^2 \Big(e^{2 T} + \f{\gamma^2 L_o^2 \delta_{\t{c}}^2}{64 m^2} e^{-2 T}\Big), \nonumber\\
\label{eq:bt}
& \cosh \big(\delta_{b} \,\t{b}(T)\big) = {b}_o \tanh\Big(\f{1}{2} {b}_o T + 2 \tanh^{-1}\frac{1}{b_o}\Big), \nonumber\\
\label{eq:pbt}
&\t{p}_b(T) = \nonumber - 2m \gamma L_o  \delta_{\t{b}} \sinh(\delta_{\t{b}} \t{b}(T))/(\gamma^2 \delta_{\t{b}}^2 - \sinh^2(\delta_{\t{b}} \t{b}(T))). \nonumber
\ea
Here $\delta_{\t{b}} = \delta_{b}$ and $\delta_{\t{c}} = \delta_{c}$ given in Eq. (\ref{db-dc}), but now, $T >0$ rather than $T<0$ (or,  $r>2m$ rather than $r<2m$). Note that there is no change in the equations in the $c$-sector, whence the Dirac observable $m$ is the same as in the interior. However, in the $b$-sector, there are some changes in signs and the trigonometric functions are replaced by the corresponding hyperbolic-trigonometric  ones. 
\smallskip

\emph{Properties of the quantum extended Kruskal space-time:} For any given value of  $m$,  
we have an effective trajectory for the `interior'  with $T_{0}< T< 0$ and one for the `exterior'  with $0< T< \infty$. It is easy to verify that these two trajectories are smooth continuations of each other at $T=0$. In the space-time picture,  although the $T,x$ chart breaks down at $T=0$ (just as the Schwarzschild  $r,t$ chart breaks down at $r=2m$), the effective 4-geometry is in fact smooth. Indeed, since the 4-metric  of Eq. (\ref{metric}) is spherically symmetric, and the product
$(g_{xx})\, (g_{TT})$ is smooth and tends to $1$  as $T \to 0$, as in GR one can introduce Eddington-Finkelstein type coordinates to make the smoothness of the 4-metric manifest. By repeating this procedure across other horizons one arrives at the smooth extension of Kruskal space-time with an infinite number of asymptotic, trapped and anti-trapped regions as depicted in Fig.  \ref{fig:4}.

Not only are the curvature scalars bounded on the full extension but, interestingly, for macroscopic black holes these bounds are uniform, i.e., independent of $m$.  As one would expect, the upper bounds are reached on the transition surface $\T$ and have the following asymptotic forms  in the large $m$ limit:
 \be C_{abcd}C^{abcd}\mid_{\T}\,\,=\,\, \frac{1024\pi^2}{3\gamma^4\Delta^2}+{\cal O}\Big(\big(\f{\Delta}{m^{2}}\big)^{\f{1}{3}}\, \ln \f{m^{2}}{\Delta}\Big) ,\ee 
%\ba
%&R_{ab}R^{ab}\mid_{\T}\,\,=\,\,\frac{256\pi^2}{\gamma^4\Delta^2}+{\cal O}\Big(\big(\f{m_{\rm Pl}}{m}\big)^{\f{2}{3}}\,\ln \f{m}{m_{\rm Pl}} \Big), \\%\nonumber\\
%&C_{abcd}C^{abcd}\mid_{\T}\,\,=\,\, \frac{1024\pi^2}{3\gamma^4\Delta^2}+{\cal O}\Big(\big(\f{m_{\rm Pl}}{m}\big)^{\f{2}{3}}\, \ln \f{m}{m_{\rm Pl}}\Big) , \ea 
%
and similarly for other curvature invariants \cite{aos}. Note that the area gap $\Delta$ appears in the denominator; finiteness of all upper bounds can thus be directly traced back to quantum geometry. In the limit $\Delta \to 0$,   the leading term diverges, reflecting the fact that in GR this invariant tends to infinity at the singularity. 

Although there is no physical matter, Einstein's vacuum equations receive quantum corrections. It is often convenient to re-interpret the non-vanishing Einstein tensor as an effective stress-energy tensor $\mathfrak{T}_{ab}$, induced by quantum geometry. As one would expect, $\mathfrak{T}_{ab}$ violates the strong energy condition and the violation is large near the transition surface ${\T}$. For example, in Planck units, for ${M}= 10^{6} $ the energy density near $\T$ is  $\rho \sim  -1 $. However,  $\mathfrak{T}_{ab}$ drops off very quickly as one moves away from the transition surface. At the horizon, $\mathfrak{T}_{ab}\mathfrak{T}^{ab} \sim 10^{-35} $. In the `exterior' asymptotic region  it further decays very rapidly \cite{aos}. Thus, as LQC, quantum geometry effects are sufficiently strong in the Planck regime to resolve the singularity but they decay rapidly for macroscopic black holes (indeed, already when  $M$ is only $10^{6}$!). For astrophysical black holes, then,  in contrast to some recent proposals (see, e.g. \cite{sg}), GR provides an excellent approximation near and outside horizons in our effective theory.

Since the effective metric is asymptotically flat, one can calculate the ADM  mass $M_{\rm ADM}$ in various asymptotic regions. A natural question is whether there is mass amplification  $A[I,III]$ as one goes from an asymptotic region I to another to its future, III (see Fig. \ref{fig:4}).  In the large $m$ limit, we find \cite{aos}:
\be \label{amplification} 
 A [I,\, III] = 1+ \mathcal {O}\Big[\Big(\f{m_{\rm Pl}}{m}\Big)^{\f{2}{3}} \,\,\ln \Big(\f{m}{m_{\rm Pl}}\Big)\Big].\ee
Thus, if the radius of the black hole horizon in asymptotic region I is, say, $r_{\rm B} =3$ km, corresponding to a solar mass, that of the white hole horizon in asymptotic region III is 
$r_{W} \approx (3 + \mathcal{O}(10^{-25}))$ km.

At first this is puzzling from considerations of the Komar mass  $K[S] := -({1}/{8\pi G})\, \oint _{S}\epsilon_{ab}{}^{cd} \nabla_{c}X_{d} \, \d S^{ab}$,  associated with the translational Killing field $X^{a} \equiv\partial/\partial x$ and a round sphere $S$.  If $S_{1}$ is chosen to lie on the BH horizon in region I, and $S_{2}$ on the white hole horizon in region III, we have: $ K[S_{2}] -K[S_{1}] = 2 \int_{\bar\Sigma} \big(\mathfrak{T}_{ab} -(\mathfrak{T}/2) g_{ab}\big) X^{a} \d V^{b}$, where $\bar\Sigma$ is a time-like 3-surface joining $S_{1}$ and $S_{2}$. Now,  the integrand in this 3-surface integral is large and negative because of properties of $\mathfrak{T}_{ab}$ in the `interior' region. Therefore one would expect $K[S_{2}]$ to be very different from $K[S_{1}]$. How can the two ADM masses be then equal? It turns out that  $\mathfrak{T}_{ab}$ is subtle:  The effective energy density is negative and sufficiently large to resolve the singularity, but also delicately balanced to make the volume integral precisely equal to $-2K[S_{1}]$ (in the large $m$ limit), making $K[S_{2}] = -K[S_{1}]$! Finally, by smoothness of the effective space-time geometry, while $X^{a}$ is future directed in region I, it is past directed in region III. Since the ADM mass refers to the unit \emph{future pointing} asymptotic time translation, $K[S_{2}] = -K[S_{1}]$ is precisely the necessary and sufficient condition for $M_{\rm ADM}(I) = M_{\rm ADM} (III)$. This calculation brings out the fact that there are highly non-trivial constraints if one wants to achieve both, singularity resolution, and (nearly) unit amplification factor for the mass in the transition from the trapped to the anti-trapped region. Our effective geometry satisfies them automatically.

\emph{Comparison:} Over the last decade, the fate of Schwarzschild singularity in quantum gravity has been discussed extensively in LQG \cite{ab,lm,ck,bv,dc,cgp,cs,oss,cctr,yks,js,dc2,bkd,gop,djs,mc,bcdhr,crfv,hhcr}. For comparison we will focus only on  the large body of work  on effective dynamics in the Schwarzschild `interior'. In all these investigations the black hole singularity is resolved. However,  subsequent analysis has shown that the detailed dynamics in these works have physically undesirable features: (i) Physical results in \cite{ab,lm,ck,cgp} can depend on the fiducial cell (through $L_{o}$), whence  details of their predictions have no invariant significance; (ii)  Approaches \cite{bv,dc,cs,oss,cctr,bkd,djs} lead to large quantum effects in low curvature regions. For example,  for large black holes, the quintessentially  quantum transition surface $\T$ can emerge in regions with arbitrarily small curvature in some approaches  (e.g. \cite{cs}), while dynamics of other approaches  \cite{bv,dc,cctr,bkd,djs} drive their effective trajectories to phase space regions where their basic underlying assumptions are violated \cite{aos}; (iii)  There can be large mass amplification in the transition from black to while hole horizon. For example, \cite{cs} leads to the amplification $A[I,\,III] \approx (m/m_{\rm Pl})^{3}$ as one evolves from the trapped region to the anti-trapped one; thus  if $r_{B} = 3$ km corresponding to solar mass, then $r_{W} \approx 10^{93}$Gpc! Physical origin of this enormous effect has remained unclear.  Our effective description is free of all these undesirable features. (For details, see sections IV D and VI  of \cite{aos}.)
 
 Another key difference from previous investigations is that  they considered only the Schwarzschild `interior' and treated it as a homogeneous (Kantowski-Sachs)  cosmology, emphasizing issues that are central to anisotropic cosmological models such as bounces of scale factors (see, e.g.,\cite{bv,dc,dc2}) and boundedness of anisotropic shears (see, e.g.,\cite{js}). By contrast, our effective theory encompasses both the interior and the asymptotic regions  and our focus is on black hole aspects that they did not consider: trapped and anti-trapped surfaces in the `interior' region, the transition surface $\T$, properties of the Komar mass and the ADM mass in the asymptotic region. Finally, transition from a trapped to an anti-trapped region also appears  in the path integral approach to LQG \cite{crfv,hhcr}. However, there one considers gravitational collapse, and the focus is on calculating transition amplitudes between specific asymptotic configurations under some approximations, while we consider the eternal Kruskal black hole and our quantum corrected equations provide a detailed description of singularity resolution within the effective theory.

Our analysis also provides a concrete context to compare and contrast singularity resolution due to quantum geometry in LQG, and an  AdS/CFT-based expectation that quantum gravity will/should not resolve certain bulk singularities, including those of  the classical  Schwarzschild-Anti-de Sitter black holes \cite{eh}.  This conclusion about the bulk geometry is arrived at  indirectly, starting from physically desirable properties of quantum field theories on the boundary, assuming the bulk-boundary duality. In LQG, on the other hand one works \emph{directly} in the bulk. There is tension between the two in that our effective theory does resolve the Schwarzschild singularity in a coherent fashion. However,  there is no contradiction since the AdS/CFT arguments do not go through in the asymptotically flat context. Therefore, it is of interest to investigate if the effective theory proposed here can be extended to the asymptotically Anti-de Sitter case. A result in either direction will provide valuable guidance.

\emph{Limitations:} We conclude with a discussion of limitations of our approach. While our quantum corrected effective geometries are of interest in their own right because of their various properties, so far, it has not been arrived at systematically starting from the full quantum theory, as was done in LQC \cite{jw,vt,asrev}. This step will likely require significant  effort because one would have to first  simplify the explicit action of our quantum Hamiltonian constraint considerably. Next, stability analysis of the extended Kruskal space-time will have to be carried out using the analog of the well-developed perturbation theory on quantum cosmological space-times (see e.g. \cite{apbook}). The most important limitation is that we have discussed eternal rather than dynamical black holes.  To address key conceptual issues such as the possibility of information loss, one would  have to consider black holes formed by gravitational collapse, where space-time structure is significantly different \cite{aa-ilqg,cdl}.  Nonetheless, just as the analysis of quantum fields on the Kruskal space-time provided useful tools to investigate the Hawking process in physically more realistic collapsing situations, techniques developed in this quantum extension of Kruskal space-time should be helpful in the analysis of the end point of  the dynamical evaporation process.\smallskip

%\section*{Acknowledgments}  
%\textbf{Acknowledgments:}  

\begin{acknowledgements} This work was supported in part by the  NSF grants PHY-1454832 and PHY-1505411, grant UN2017-9945 from the Urania Stott Fund of the Pittsburgh Foundation, the Eberly research funds of Penn State, and by Project. No. MINECO FIS2014-54800-C2-2-P from Spain and its continuation Project. No. MINECO FIS2017-86497-C2-2-P.  We thank Eugenio Bianchi and Hal Haggard for discussions. 
\end{acknowledgements}

\end{document}